\documentclass{article}
\usepackage{amssymb}
\usepackage{amsfonts}
\usepackage{amsmath}

\newtheorem{theorem}{Theorem}
\newtheorem{acknowledgement}[theorem]{Acknowledgement}

\input{tcilatex}
\begin{document}

\title{\textbf{A} \textbf{causal viscous cosmology without singularities}}
\author{Carlos E. Laciana \\
Departamento de Hidr\'{a}ulica, Facultad de Ingenier\'{\i}a, \\
Universidad de Buenos Aires, \\
Av. Las Heras 2214, Ciudad Aut\'{o}noma de Buenos Aires, \\
C1127AAR, \ Argentina.\\
email: clacian@fi.uba.ar
\ }
\date{ \ \ }
\maketitle

\begin{abstract}
An isotropic and homogeneous cosmological model with a source of dark energy
is studied. That source is simulated with a viscous relativistic fluid with
minimal causal correction. In this model the restrictions on the parameters
coming from the following conditions are analized: a) energy density without
singularities along time, b) scale factor increasing with time, c) universe
accelerated at present time, d) state equation for dark energy with
\textquotedblleft $w$\textquotedblright\ bounded and close to -1. It is
found that those conditions are satified for the following two cases. i)
When the transport coefficient ($\tau _{\Pi }$), associated to the causal
correction, is negative, with the aditional restriction $\zeta \left\vert
\tau _{\Pi }\right\vert >2/3$, where $\zeta $ is the relativistic bulk
viscosity coefficient. The state equation is in the \textquotedblleft
phantom\textquotedblright\ energy sector. ii) For $\tau _{\Pi }$ positive,
in the \textquotedblleft k-essence\textquotedblright\ sector. It is
performed an exact calculation for the case where the equation of state is
constant, finding that option (ii) is favored in relation to (i), because in
(ii) the entropy is always increasing, while this does no happen in (i).
\end{abstract}

\section{Introduction}

As stressed in ref. \cite{Sahni}, there is relevant experimental evidence
about the acceleration of the universe. That effect can be attributed to the
so-called dark energy, which can be interpreted as a term of "negative
pressure" in the Einstein equations, as remarked in ref. \cite{Padmanabhan}.

The most cited experimental observations related to universe acceleration
are due to Perlmutter et al. \cite{Perlmutter} and Riess et al. \cite{Riess}%
, which include measures of the redshift of supernovas. The results are
compatible with the addition of the cosmological constant to the Einstein
equations. This cosmological constant would be analogous to a negative
pressure that leads the universe to an accelerated expansion.

The dark energy can be described by the state equation $\mathit{p}=\mathit{w}%
\rho $, with $w<-1/3$, in order to have an accelerated universe \cite%
{Caldwell}. In agreement with experimental observations \cite{Nakamura}, the
value of $w$ would be very close to $-1$ ($w=-1.04_{-0.10}^{+0.09}$). From a
theoretical point of view, the different values of $w$ considered in the
literature come from some Lagrangian formulations of field theories (for a
review, see ref. \cite{Copeland}). It is so for a field with minimal
coupling, i.e., $-1\leq w\leq 1$, which is known as "quintessence". Also for
potentials coming from string theory as, for example, the formulation known
as "k-essence", with $-1\leq w\leq -1/3$. From s-brane, in superstring
theory, it is obtained for the "phantom" field $w<-1$. The last case
corresponds to a minimal coupling field, but with contrary sign in the
cinetic term with respect to an ordinary field. This model of dark energy
shows an anomalous behavior of difficult justification \cite{Nojiri}, for
example, it leads to a singularity in the density of energy at finite time,
known as "Big Rip" \cite{Nojiri 2004}; moreover, it describes energy density
as an increasing function of time, with the consequence of the violation of
the dominant energy condition \cite{Hawking and Ellis} and, in addition, it
leads to negative values of entropy.

Another used approach is based on considering a relativistic viscous fluid
as a matter source of the Einstein equations. For example, in ref. \cite%
{Fabris} the matter of the universe is a viscous fluid; moreover, the
approximation introduced originally by Eckart \cite{Eckart} and later
improved by Landau-Lifshitz (L-L) \cite{Landau F} is used. When this model
is applied to an isotropic and homogeneous universe, the acceleration in the
expansion is produced by the bulk viscosity. The works \cite{Murphy} and 
\cite{Zimdahl} follow this line of research. More recently, ref. \cite%
{Brevik 2013} shows that, if the bulk viscosity coefficient is negative,
then the expansion decreases. On the other hand, ref. \cite{Lima} proves the
equivalence between the dynamics coming from a macroscopic approach based on
the bulk viscosity term and a particle creation model that describes the
phase of a slow-roller inflation.

Nevertheless, the application of the causal approach in the matter source is
usually not found in the literature. As it is well known, the L-L theory has
problems of stability and admits propagation of superluminal signals \cite%
{Hiscock}. This last fact, as shown in papers \cite{Dolgov} and \cite%
{Liberati}, implies violations of causality. In order to avoid causality
violation problems, some theories of viscous fluid, that incorporate second
order gradients in the velocities, have been developed \cite{Israel-Stewart}%
, \cite{Romatschke IJMP}.

However, when the purpose is the study of the dynamics of the universe, the
theories with causal correction give very complex expressions for the
energy-momentum tensor. It is for that reason that, in the present work, a
much simpler approach is proposed, in line with ref. \cite{Koide}, where the
anomalies of the L-L theory are avoided with a minimum of complexity.

In ref. \cite{Koide} it is shown how the parabolic equation for the
perturbation in the velocities, with the causality violation problem, can be
turned into an hyperbolic equation without that problem. In order to give a
causal behaviour, a time delay in the propagation of information, which the
authors call "relaxation time", is introduced. This is implemented by means
of a memory function (Green function). This formulation introduces a
correction in the bulk viscosity, in agreement with the correction coming
from the Israel-Stewart theory \cite{Israel-Stewart}, or from the more
general proposal of ref. \cite{Romatschke CQG} or \cite{Maartens} at the low
order, necessary to avoid anomalies with causality.

The aim of this work is to perform an analysis of the cosmological equations
for a viscous fluid model, in order to understand the importance of the
causal correction in the dynamics of the universe. The purpose is also to
determine the influence of the correction term mentioned above over the
entropy of the universe.

With this aim, the present article is organized as follows. In the next
section, the cosmological model and the dynamical equations resulting from
the causal viscous correction are introduced. Section 3 is devoted to the
analisis of the restrictions on the parameters. In Section 4, the
contribution of the causal corrections to entropy is studied. Finally, in
Section 5 the main conclusions are presented.

\section{The cosmological model}

Let us consider an isotropic and homogeneous universe with matter modelled
as a relativistic fluid. Then, a spatially flat Friedmann-Robertson-Walker
(FRW) universe is considered. From here on, the units for which $c=8\pi
G=k=1 $ and the signature for the metrics $(+,-,-,-)$ are used, as in ref. 
\cite{Landau C}.

\subsection{Dynamic equations}

The dynamics will be described by the Einstein equations with the
energy-momentum tensor (EMT) as a source of viscous fluid (see ref. \cite%
{Landau C} or \cite{Gravitation}), i.e.:

\begin{equation}
R_{\mu \nu }-\frac{1}{2}g_{\mu \nu }R=T_{\mu \nu },  \label{2.1}
\end{equation}%
where de EMT can be factorized as \cite{Romatschke CQG}:

\begin{equation}
T_{\mu \nu }=\widetilde{T}_{\mu \nu }+\Pi _{\mu \nu }.  \label{2.2}
\end{equation}%
$\widetilde{T}_{\mu \nu }$ is the ideal fluid part, given by

\begin{equation}
\widetilde{T}_{\mu \nu }=\left( p+\rho \right) u_{\mu }u_{\nu }-pg_{\mu \nu
},  \label{2.3}
\end{equation}%
where $\rho $ is the density of energy (dark energy in our case), $p$ is the
pressure, $u_{\mu }$ the four-velocity, and $g_{\mu \nu }$ is the space-time
metrics. Given the convention used here, we have the normalization equation $%
u_{\mu }u^{\mu }=1.$

The tensor $\Pi _{\mu \nu }$ in Eq.(\ref{2.2}) is the viscosity term. This
term can also be factorized into two tensors, one of them the traceless part
($\pi _{\mu \nu }$ ), related with the shear viscosity, and the other part
with non-vanishing trace ( $\Pi $), representing the bulk viscosity. Then,
we have (as in \cite{Romatschke IJMP})

\begin{equation}
\Pi _{\mu \nu }=\pi _{\mu \nu }+\Delta _{\mu \nu }\Pi ,  \label{2.4}
\end{equation}%
with

\begin{equation}
\Delta _{\mu \nu }=g_{\mu \nu }-u_{\mu }u_{\nu }.  \label{2.5}
\end{equation}

It is convenient to re-write $\widetilde{T}_{\mu \nu }$ in the form:

\begin{equation}
\widetilde{T}_{\mu \nu }=\rho u_{\mu }u_{\nu }-p\Delta _{\mu \nu }.
\label{2.6}
\end{equation}%
Then,

\begin{equation}
\widetilde{T}_{\mu }^{\mu }=\rho -3p.  \label{2.7}
\end{equation}%
By replacing Eq.(\ref{2.4}) into Eq.(\ref{2.2}), we obtain

\begin{equation}
T_{\mu \nu }=\rho u_{\mu }u_{\nu }-\left( p-\Pi \right) \Delta _{\mu \nu
}+\pi _{\mu \nu }.  \label{2.8}
\end{equation}%
Therefore, now we have

\begin{equation}
T_{\mu }^{\mu }=\rho -3\left( p-\Pi \right) ,  \label{2.9}
\end{equation}%
because

\begin{equation}
\pi _{\mu }^{\mu }=0.  \label{2.10}
\end{equation}%
By comparing Eq. (\ref{2.7}) with Eq. (\ref{2.9}), we can see that the
quantity $p-\Pi $ is equivalent to a corrected pressure. Then, we can say
that $-\Pi $ is a kind of \textquotedblleft negative
pressure\textquotedblright\ that represents the viscosity effect. On this
basis, it is convenient to define

\begin{equation}
p^{\dagger }\equiv p-\Pi  \label{2.11}
\end{equation}

By taking the trace in Eq. (\ref{2.2}), and by using Eq. (\ref{2.9}) with
the definition (\ref{2.11}) and the Hubble coefficient $H=\overset{\cdot }{a}%
/a$, with $a(t)$ the scale factor, one finds

\begin{equation}
\overset{\cdot }{H}+2H^{2}=\frac{1}{6}\left( \rho -3p^{\dagger }\right) .
\label{2.12}
\end{equation}%
If the 00 component is taken from Eq. (\ref{2.1}), and the term $\pi _{\mu
\nu }$ is neglected in Eq. (\ref{2.8}) because at lower order it goes as $%
\partial _{<\alpha }u_{\beta >}$ (see ref. \cite{Romatschke IJMP}), then one
obtains

\begin{equation}
H^{2}=\frac{1}{3}\rho .  \label{2.13}
\end{equation}%
For the closed universe, a term of the form $a^{-2}$ must be added in Eqs. (%
\ref{2.12}) and (\ref{2.13}). Then, when $a>>1$, the formulation is
approximately that corresponding to a flat universe. By derivation of Eq. (%
\ref{2.13}), and by replacing the result in Eq. (\ref{2.12}), the following
well-known and useful expression can be obtained:

\begin{equation}
\overset{\cdot }{\rho }+3H\left( p^{\dagger }+\rho \right) =0.  \label{2.14}
\end{equation}%
It is easy to prove that the last equation is valid for both flat and closed
universe in exact way.

\subsection{Causal correction}

The functional form of the quantity $\Pi $ was developed in the references 
\cite{Israel}, \cite{Israel-Stewart}, \cite{Israel-Stewart PL}, \cite%
{Maartens}, and the generalization to second order in velocity gradients is
also given by \cite{Romatschke CQG}. Hereafter, the approach by Koide et al. 
\cite{Koide} will be used. Such proposal corresponds to the lowest order, in
which the violation of causality does not occur. The quantity $\Pi $, in
that reference, is approximated by:

\begin{equation}
\Pi \simeq \zeta \triangledown _{\mu }u^{\mu }-\tau _{\Pi }u^{\mu
}\triangledown _{\mu }\Pi ,  \label{2.15}
\end{equation}%
where $\zeta $ is the bulk viscosity coefficient and $\tau _{\Pi }$,
sometimes called "second viscosity coefficient", comes from the causal
correction (see \cite{Romatschke CQG}). The product $\zeta \tau _{\Pi
}\equiv \tau _{R}$ is called \textquotedblleft relaxation
time\textquotedblright\ (see \cite{Koide}).

As we can see, Eq. (\ref{2.15}) is an implicit equation of $\Pi $;
therefore, to calculate the source of the dynamical equation it is useless.
However, we can deduce an approximate, but explicit, expression to determine 
$\Pi $. In order to do this, we can substitute Eq. (\ref{2.15}) into itself
and neglect the terms higher than the first order in $\tau _{\Pi }$. Then,
one obtains

\begin{equation}
\Pi \simeq \zeta \triangledown _{\mu }u^{\mu }-\zeta \tau _{\Pi }u^{\mu
}\triangledown _{\mu }\triangledown _{\alpha }u^{\alpha }.  \label{2.16}
\end{equation}%
The first term of Eq. (\ref{2.16}) corresponds to the Landau-Liftshitz
theory \cite{Landau F}, while the second term is the minimun necessary in
order to avoid causality violation.

Now we will express $\Pi $ as a function of the scale factor $a(t)$ and its
derivatives. So, it is convenient to employ the continuity equation ($%
\triangledown _{\mu }\left( nu^{\mu }\right) =0$ with $n$ the density
number) and to express the equations by means of the proper time. So, we
finally get

\begin{equation}
\Pi =3\zeta H-3\zeta \tau _{\Pi }\overset{\cdot }{H}.  \label{2.17}
\end{equation}%
This functional form for $\Pi $ is particularly convenient to solve the
dynamical equation.

\subsection{System of equations to be solved}

The equation of state (EoS), which relates the pressure and the density of
dark energy, is added to the equations given above. Following ref. \cite%
{Frampton}, we can use the following expression for the dark energy

\begin{equation}
w=\frac{p}{\rho }=-1-\lambda \rho ^{\alpha -1},  \label{2.18}
\end{equation}%
with $\alpha $ an arbitrary parameter.

On the other hand, by replacing Eq. (\ref{2.14}) into Eqs. (\ref{2.13}) and (%
\ref{2.18}) we obtain

\begin{equation}
\left( 1+\frac{3}{2}\zeta \tau _{\Pi }\right) \overset{\cdot }{H}-\frac{%
3^{\alpha }}{2}\lambda H^{2\alpha }-\frac{3}{2}\zeta H=0.  \label{2.19}
\end{equation}%
As we will see, by the resolution of Eq. (\ref{2.19}), in some cases it is
possible to obtain an exact expression for the dark energy density as a time
dependent function.

The questions that we want to answer in this paper are referred to the
conditions that the parameters must satisfy for the following requirements
be fulfilled: i) energy density without singularities at finite time, ii)
scale factor $a(t)$ as an increasing function of the time, iii) accelerated
universe at present time, and iv) $w$ close to $-1$.

By operating with Eqs. (\ref{2.12})-(\ref{2.18}), it is easy to obtain the
following set of equations, useful to test the above requirements:

\begin{equation}
\Delta t=\frac{1}{\sqrt{3}}\int_{\rho _{P}}^{\rho }\frac{M\left( \rho
\right) }{\rho ^{1/2}}d\rho ,  \label{2.20}
\end{equation}

\begin{equation}
\ln \left( \frac{a}{a_{P}}\right) ^{3}=\int_{\rho _{P}}^{\rho }M\left( \rho
\right) d\rho ,  \label{2.21}
\end{equation}

\begin{equation}
\frac{\overset{\cdot \cdot }{a}}{a}=\frac{1}{3}\rho \left( 1+\frac{3}{2}%
\frac{1}{\rho M\left( \rho \right) }\right) ,  \label{2.22}
\end{equation}%
with $\Delta t\equiv t-t_{P}$ (subindex \textquotedblleft $P$
\textquotedblright\ indicate Planck era), and

\begin{equation}
M\left( \rho \right) \equiv \left( 1+\frac{3}{2}\zeta \tau _{\Pi }\right)
/\left( \lambda \rho ^{\alpha }+\sqrt{3}\zeta \rho ^{1/2}\right) .
\label{2.23}
\end{equation}

Eqs. (\ref{2.20}) - (\ref{2.23}), plus Eq. (\ref{2.18}), are the set of
equations that, in the next section, will be used to analyze the
restrictions in the parameters necessary to avoid singularities in the
physical quantities. In particular, the main interest is to determine the
influence of the causal correction on the results.

\section{System of equations to be solved}

In this section, the conditions that the quantities on the left side of Eqs.
(\ref{2.20}) - (\ref{2.23}) and the state equation (\ref{2.18}) must satisfy
will be formulated, i.e.:

\begin{description}
\item i) The condition on $\Delta t=\Delta t\left( \rho \right) $. In this
case, two possibilities are considered:

\begin{description}
\item I) The dark energy density as a decreasing function of time, i.e.

\item $%
\begin{array}{c}
\lim \\ 
\rho \rightarrow \rho _{f}%
\end{array}%
\Delta t=\infty $, for $\rho _{f}<\rho _{P}$, (in particular, $\rho _{f}$
can be zero).

\item II) The dark energy density as an increasing function of time, i.e.

\item $%
\begin{array}{c}
\lim \\ 
\rho \rightarrow \rho _{f}%
\end{array}%
\Delta t=\infty $, for $\rho _{f}>\rho _{P}$. A particular case is $\rho
_{f}=\infty $, which is known in the literature as "Little Rip" \cite%
{Frampton}.
\end{description}

\item ii) The scale factor $a(t)$ as an increasing function of time, i.e.

\item $%
\begin{array}{c}
\lim \\ 
\rho \rightarrow \rho _{f}%
\end{array}%
a\left( \rho \right) =a_{f}$, such that $a_{f}>a_{o}$, where $a_{o}$ is the
scale factor observed at present .

\item iii) Accelerated universe (at least for $t\sim t_{o})$, i.e. $\overset{%
\cdot \cdot }{a}$ $>0$.

\item iv) $w\sim -1\pm 0.1$
\end{description}

\subsection{General restrictions}

In this subsection, the restrictions on the parameters due to the above
conditions are analyzed in general. In particular, the two possibilities I
and II are considered:

\subsubsection{I) $\protect\rho $ as a decreasing function}

Condition (i) tells us that $\Delta t$ is a decreasing function of $\rho $,
because $\rho $ decreases with $t$. Then, the following inequality must be
satisfied: $d\Delta t/d\rho <0$. As a consequence, from Eq. (\ref{2.20}) the
implication is

\begin{equation}
M(\rho )<0.  \label{2.24}
\end{equation}

From condition (ii), $\ln \left( \frac{a}{a_{P}}\right) ^{3}$ is a
decreasing function of $\rho $. Therefore, the inequality (\ref{2.24}) is
implied again, i.e., condition (ii) does not introduce a new restriction. It
is noteworthy that the condition (\ref{2.24}) may be satisfied by negative
numerator or denominator of $M(\rho )$. In the first case, it should be $%
\tau _{\Pi }$ $<0$, with the additional condition 
\begin{equation}
\zeta \left\vert \tau _{\Pi }\right\vert >2/3.  \label{2.25}
\end{equation}%
In the second case, we should make $\lambda <0$ (quintessence sector,
k-essence, tachyon field, etc.), with the additional condition

\begin{equation}
\left\vert \lambda \right\vert >\sqrt{3}\zeta \rho ^{\frac{1}{2}-\alpha }.
\label{2.26}
\end{equation}%
Since in model I $\rho \left( t\right) $ decreases with time, the above
condition requires $\alpha \leq 1/2$ in order to avoid the indefinite
increase of the boundary of $\left\vert \lambda \right\vert $, or that $\rho
\left( t\right) $ tends to a finite value, as we shall see in the next
subsection.

From condition (iii), the following inequality (besides that of (\ref{2.24}%
)) must be satisfied:

\begin{equation}
\rho >\frac{3}{2}\left\vert M\left( \rho \right) \right\vert ^{-1},\forall
\rho \leq \rho _{P}.  \label{2.27}
\end{equation}%
This implies that there is a density value, let's call it $\rho _{c}$, below
which the condition (iii) is not met. This value will depend on how close to 
$1$ the value of $\frac{3}{2}\zeta \left\vert \tau _{\Pi }\right\vert $ is,
and how small $\lambda $ is. This will be clear when, in the next
subsection, a specific case will be analyzed.

Condition (iv) implies that

\begin{equation}
\left\vert \lambda \right\vert \rho ^{\alpha -1}\lesssim 0.1.  \label{2.28}
\end{equation}%
Then, for $w$ limited, it must be $\alpha \geqslant 1$, which is in
contradiction with the condition coming from (\ref{2.24}).

\subsubsection{II) $\protect\rho $ as an increasing function}

Condition (i) in this case tells us that $\Delta t$ is an increasing
function of $\rho $, i.e.: $d\Delta t/d\rho >0$. Therefore,

\begin{equation}
M(\rho )>0.  \label{2.29}
\end{equation}

Condition (ii) $\Rightarrow d\ln (a/a_{p})/d\rho >0\Rightarrow $(\ref{2.29}).

Condition (iii) is directly satisfied.

Condition (iv) is the same as in the case of (\ref{2.28}), but with $\lambda
=\left\vert \lambda \right\vert $ (phantom energy sector). But, since $\rho $
can grow indefinitely, in order to keep $w$ bounded, in this case it should
hold that $\alpha \leq 1$.\bigskip

So, as we see, both cases (I and II) share the same condition $\alpha =1$.
Hence, it is interesting to analyze this case in more detail, a task that we
will undertake in the next subsection.

\subsection{The particular case $\protect\alpha =1$}

Why to study the detail of a particular case, if the general conditions were
just given? The reason is that the general conditions are valid for
functions with monotonic behavior, i.e., functions that increase or decrease
along all the time interval. However, this analysis is beyond that case. For
example, $\overset{\cdot }{a}(t)$ is an increasing function of time in a
range of time, but in another range it is decreasing function, in such a way
that in the present time the universe is accelerated, consistent with
observations, but in a far later time, slowdown occurs. Then, this example
is not covered by the criterion given above. Moreover, this case satisfies
automatically one of the required conditions: the EoS remains bounded during
the whole evolution of the universe.

Then we start with Eq. (\ref{2.19}), which is solved exactly by means of the
methodology of ref. \cite{Brevik 2013}, we obtain the following solution:

\begin{equation}
H(t)=-\frac{\zeta }{2\lambda }\left[ 1+\coth \left( \gamma t\right) \right] ,
\label{2.30}
\end{equation}%
with $\gamma \equiv \left( 3\zeta /4\right) /\left( 1+3\zeta \tau _{\Pi
}/2\right) .$

The functional form of this equation shows us, that for any natural number $%
n $, $d^{n}H(t)/dt^{n}\neq \infty $ $\forall $ $t\neq 0$ then, the studied
case, does not present the singularity, identified as Type IV in the
literature \cite{JLS-GS}, \cite{NoOdiOik}.

Taking into account that $H=d\ln a/dt$, we can integrate in time to obtain $%
a=a\left( t\right) $, which results:

\begin{equation}
a(t)=a_{P}\left[ \frac{1-e^{2\gamma t}}{1-e^{2\gamma t_{P}}}\right] ^{-\zeta
/2\lambda \gamma }.  \label{2.31}
\end{equation}%
From Eq. (\ref{2.31}) we can see that, for $a(t)$ be an increasing function
of $t$ , as $\zeta >0$ (see ref. \cite{Landau F}), there are the following
possibilities:

\subsubsection{Phantom energy sector: $\protect\lambda >0$ $.$}

\paragraph{a) $\protect\gamma <0$}

Then $\tau _{\Pi }$ $<0$ with the condition (\ref{2.25}) is required in
order to satisfy the inequality on $\gamma $.

Then, we can write

\begin{equation}
a(t)=a_{P}\left[ \frac{1-e^{-2\left\vert \gamma \right\vert t}}{%
1-e^{-2\left\vert \gamma \right\vert t_{P}}}\right] ^{\zeta /2\lambda
\left\vert \gamma \right\vert },  \label{2.32}
\end{equation}%
whereupon

\begin{equation}
\begin{array}{c}
\lim \\ 
t\rightarrow \infty%
\end{array}%
a(t)=a_{P}\left[ 1-e^{-2\left\vert \gamma \right\vert t_{P}}\right] ^{-\zeta
/2\lambda \left\vert \gamma \right\vert },  \label{2.33}
\end{equation}%
As we can see from Eq. (\ref{2.33}), $a\left( \infty \right) \neq \infty $,
but it can be as large as we want. In fact, if for example $\left\vert
\gamma \right\vert \sim 1$, as $t_{P}\ll 1$, developing the exponential up
to linear term, we obtain $a\left( \infty \right) \simeq \left(
2t_{P}\right) ^{-\zeta /2\lambda }$: also the exponent can be in absolute
value as large as we want, if $\lambda \ll 1.$The last requirement is
consistent with the fact that $w\sim -1$ in the phantom energy region.

Moreover, by using Eq.(\ref{2.30}) and Eq. (\ref{2.13}), the energy density
results

\begin{equation}
\rho (t)=\frac{3}{4}\left( \frac{\zeta }{\lambda }\right) ^{2}\left[ 1-\coth
\left( \left\vert \gamma \right\vert t\right) \right] ^{2},  \label{2.34}
\end{equation}%
Clearly $%
\begin{array}{c}
\lim \\ 
t\rightarrow \infty%
\end{array}%
\rho (t)=0$ (is included in the family of models labeled by I). This result
creates some conceptual conflict in relation with the above result: if the
scale factor reaches a finite value in infinite time when the density is
null, what happened with dark energy? Did it disappear? Nevertheless, as we
saw above, this would not have a \textquotedblleft
noticeable\textquotedblright\ effect because $a(\infty )$ would be a power
of the inverse of Planck time as large as we would like.

Now we can analyze if this model gives us an accelerated universe. If the
two derivatives of Eq. (\ref{2.32}) are performed, we obtain:

\begin{equation}
\overset{\cdot \cdot }{a}\left( t\right) =\frac{2\zeta \left\vert \gamma
\right\vert }{\lambda }Q\left( t_{P}\right) \left( 1-e^{-2\left\vert \gamma
\right\vert t}\right) ^{\frac{\zeta }{2\lambda \left\vert \gamma \right\vert 
}-2}e^{^{-4\left\vert \gamma \right\vert t}}\left( \frac{\zeta }{2\lambda
\left\vert \gamma \right\vert }-e^{2\left\vert \gamma \right\vert t}\right) ,
\label{2.35}
\end{equation}%
with $Q\left( t_{P}\right) \equiv a_{P}\left( 1-e^{-2\left\vert \gamma
\right\vert t_{P}}\right) ^{-\frac{\zeta }{2\lambda \left\vert \gamma
\right\vert }}$. So, for $\overset{\cdot \cdot }{a}\left( t\right) >0$, the
following inequality must hold:

\begin{equation}
\frac{\zeta }{2\lambda \left\vert \gamma \right\vert }>e^{2\left\vert \gamma
\right\vert t}.  \label{2.36}
\end{equation}%
The two quantities of the above inequality would be equal for a
\textquotedblleft change time\textquotedblright\ $t_{c}$ given by

\begin{equation}
t_{c}=\ln \left( \frac{\zeta \left\vert \tau _{\Pi }\right\vert -2/3}{%
\lambda }\right) ^{\frac{^{\zeta \left\vert \tau _{\Pi }\right\vert -2/3}}{%
\zeta }}.  \label{2.37}
\end{equation}%
Then, the value of $\lambda $ can be set so the universe will slow at $%
t_{c}>t_{a}$, where $t_{a}$ is the current observation time. Again, we
obtain $\lambda <<1$, whereby the value of $w$ should be very close to $-1$,
in agreement with the observation.

\paragraph{b) $\protect\gamma >0$}

As is easily seen, Eq. (\ref{2.31}) gives us, in this case, a decreasing
evolution, so this possibility is discarded.

\subsubsection{K-essence sector: $\protect\lambda <0$}

\paragraph{a) $\protect\gamma >0$}

Now we start with

\begin{equation}
a(t)=a_{P}\left[ \frac{e^{2\gamma t}-1}{e^{2\gamma t_{P}}-1}\right] ^{\zeta
/2\left\vert \lambda \right\vert \gamma }.  \label{2.38}
\end{equation}%
Then $%
\begin{array}{c}
\lim \\ 
t\rightarrow \infty%
\end{array}%
a(t)=\infty $. The energy density is in this case is

\begin{equation}
\rho (t)=\frac{3}{4}\left( \frac{\zeta }{\lambda }\right) ^{2}\left[ 1+\coth
\left( \gamma t\right) \right] ^{2}.  \label{2.39}
\end{equation}%
As we see from Eq.(\ref{2.39}), a function that converges to a finite value
is obtained, i.e.: $%
\begin{array}{c}
\lim \\ 
t\rightarrow \infty%
\end{array}%
\rho (t)=3\left( \zeta /\lambda \right) ^{2}$. It is worth to notice that,
for a very small value of $\lambda $, i.e., $\left\vert \lambda \right\vert $
$<\sqrt{3}\zeta /\rho _{P}^{1/2}$, the density could grow beyond the density
at the Planck time; in that case, the model would be included in the family
that we called II. We can say that, density of dark energy will grow up to a
finite value and therefore does not become a "Little Rip" type singularity 
\cite{Frampton},\cite{BENO} for any time, provided $\lambda $ $\neq 0$.

Now, from Eq. (\ref{2.38}) the acceleration can be computed, and the
following expression is obtained:

\begin{equation}
\overset{\cdot \cdot }{a}\left( t\right) =f\left( t\right) \left( \frac{%
\zeta }{2\left\vert \lambda \right\vert \gamma }-e^{-2\gamma t}\right) ,
\label{2.40}
\end{equation}%
where $f(t)>0$ $\ \forall $ $t$. Therefore, the condition for accelerated
expansion is obtained from the condition that the quantity into the
parenthesis in Eq. (\ref{2.40}) be positive. As a consequence, the following
condition on time results:

\begin{equation}
t>\frac{1}{\zeta }\left( \frac{2}{3}+\zeta \tau _{\Pi }\right) \ln \frac{%
\left\vert \lambda \right\vert }{\frac{2}{3}+\zeta \tau _{\Pi }}.
\label{2.41}
\end{equation}%
This means that the condition $\left\vert \lambda \right\vert <\frac{2}{3}%
+\zeta \tau _{\Pi }$, is sufficient for an accelerated regimen during all
the time interval.

\paragraph{b) $\protect\gamma <0$}

In order to complete the analysis, it can stressed that the case in which
both coefficients, $\lambda $ and $\gamma $ $(\tau _{\Pi }<0)$, are negative
does not give a reasonable result, because it leads to a scale factor
decreasing over time.

\section{Causal viscosity contribution to the change of entropy}

In this section, the following question will be addressed: which is the
contribution to the entropy due to the causal corrective term in the
viscosity?

From Eq. (\ref{2.12}) we can see that the quantity $p^{\dagger }$ is
equivalent to an \textquotedblleft effective pressure\textquotedblright . On
the other hand, we can write the Gibbs equation in the form

\begin{equation}
dE=-\left( p-T\frac{dS}{dV}\right) dV.  \label{2.42}
\end{equation}%
The second term in the parenthesis can be interpreted as the heat per unit
of volume absorbed by the system as the result of viscosity. It can be
conceived as a negative pressure, analogous to the correction to the
pressure introduced in Eq. (\ref{2.11}), which we identified with $-\Pi $,
associated to the bulk viscosity. Then, it is natural to propose the
identification

\begin{equation}
T\frac{dS}{dV}\equiv \Pi ,  \label{2.43}
\end{equation}%
with $\Pi $ given by Eq. (\ref{2.17}). Then, there are two viscosity
contributions to the change of entropy, indicated as

\begin{equation}
dS=dS_{1}+dS_{2},  \label{2.44}
\end{equation}%
with

\begin{equation}
dS_{1}=\frac{3}{T}\zeta HdV,  \label{2.45}
\end{equation}

\begin{equation}
dS_{2}=-\frac{3}{T}\zeta \tau _{\Pi }\overset{\cdot }{H}dV,  \label{2.46}
\end{equation}%
Eq. (\ref{2.45}) gives the bulk viscosity contribution, and Eq. (\ref{2.46})
supplies the second bulk viscosity contribution, related with the causal
correction.

The first term of Eq. (\ref{2.44}) is always positive, since we assume that,
as in most physical systems, $T>0$. According to the theory of fluids \cite%
{Landau F}, it should hold that $\zeta >0$, since an expansion stage with $%
dV>0$ and $\overset{\cdot }{a}$ $>0$ is considered. Also in the contraction
phase the product $HdV$ remains positive. Therefore, the first term of Eq. (%
\ref{2.44}) does not break time invariance (see also ref. \cite%
{Castagnino-Laciana} for the case where particle creation is considered).

The analysis of the influence due to the second term of Eq. (\ref{2.44}) is
easier when the equation is rewritten in the following convenient form:

\begin{equation}
dS=dS_{1}\left[ 1+\tau _{\Pi }\left( \frac{\overset{\cdot }{a}}{a}-\frac{%
\overset{\cdot \cdot }{a}}{\overset{\cdot }{a}}\right) \right] .
\label{2.47}
\end{equation}%
A first remark is that the term into the parentheses that multiplies $\tau
_{\Pi }$ becomes zero for an evolution of the form $a(t)\propto \exp \left(
\alpha t\right) $. As we did above, we will separate the analysis of the
contribution to $dS$, according $\tau _{\Pi }$ is greater or smaller than
zero.

We see that, when $\tau _{\Pi }>0$, the suficient condition for the term of
entropy associated with the coefficient $\tau _{\Pi }$ gives a positive
contribution is

\begin{equation}
\frac{\overset{\cdot }{a}}{a}>\frac{\overset{\cdot \cdot }{a}}{\overset{%
\cdot }{a}}.  \label{2.48}
\end{equation}%
It has to be noticed that a solution of the form $a(t)\propto t^{\beta }$
satisfies the above condition for any $\beta $. Another possibility would be 
$\overset{\cdot \cdot }{a}/\overset{\cdot }{a}$ $<0$, but this does not
agree with observations since it would lead to a deceleration of the
universe.

When $\tau _{\Pi }<0$, the sufficient condition is

\begin{equation}
\frac{\overset{\cdot }{a}}{a}<\frac{\overset{\cdot \cdot }{a}}{\overset{%
\cdot }{a}}.  \label{2.49}
\end{equation}%
Since $\overset{\cdot }{a}$ and $a$ are positive, this condition also
implies that $\overset{\cdot \cdot }{a}$ $>0$.

\subsection{Analysis for case $\protect\alpha =1$}

To analyze the details of an exact calculation example, we can see again the
case in which the state equation is constant (i.e. with $\alpha =1$). in
order to simplify notation, it is convenient to define

\begin{equation}
\delta \equiv \frac{\overset{\cdot }{a}}{a}-\frac{\overset{\cdot \cdot }{a}}{%
\overset{\cdot }{a}}.  \label{2.50}
\end{equation}

Two subclasses are considered:

\subsubsection{a) K-essence sector: $\protect\lambda <0$ and $\protect\tau %
_{\Pi }>0.$}

By performing the derivatives of Eq. (\ref{2.38}), we can calculate $\delta $%
:

\begin{equation}
\delta =2\gamma /\left( e^{2\gamma t}-1\right) ,  \label{2.51}
\end{equation}%
with $\gamma $ defined as in Eq. (\ref{2.30}). Then, $\delta >0$ $\forall $ $%
t$ and, therefore, $dS>0$ $\forall $ $t$ . We can see also that $%
\begin{array}{c}
\lim \\ 
t\rightarrow \infty%
\end{array}%
dS=dS_{1}$. This means that, when high values of $t$, and hence of $a(t)$,
are reached, the effect due to the causal correction becomes negligible.

\subsubsection{b) Phantom energy sector: $\protect\lambda >0$ and $\protect%
\tau _{\Pi }<0.$}

Now we derive Eq. (\ref{2.32}) to calculate $\delta $. In this case we obtain

\begin{equation}
\delta =2\left\vert \gamma \right\vert /\left( 1-e^{-2\left\vert \gamma
\right\vert t}\right) .  \label{2.52}
\end{equation}%
Therefore $\delta >0$ $\forall $ $t$. However now $dS$ is

\begin{equation}
dS=dS_{1}\left( 1-\left\vert \tau _{\Pi }\right\vert \delta \right) .
\label{2.53}
\end{equation}%
If we also consider that $\zeta \left\vert \tau _{\Pi }\right\vert >2/3$ (a
restriction necessary to meet the conditions (i) - (iv)), then there is no $%
t>0$ for which entropy increases. Moreover, $%
\begin{array}{c}
\lim \\ 
t\rightarrow \infty%
\end{array}%
dS=-dS_{1}/\left( \frac{3}{2}\zeta \left\vert \tau _{\Pi }\right\vert
-1\right) $. These are then good arguments against this case.

\section{Conclusions}

A cosmological model as an isotropic and homogeneous universe, with a source
of matter that simulates dark energy, was proposed. The source consists of a
relativistic viscous fluid with minimal causal correction. Due to the
symmetry of the model, the only viscous contribution is the bulk viscosity,
which provides the negative pressure necessary to maintain an accelerated
expansion, consistent with the observations.

The constrains in the model parameters due to the following conditions, were
studied: i) energy density tending to a finite value along the time, ii)
scale factor $a(t)$ increasing function of time, iii) accelerated present
universe, and iv) state equation for dark energy $p/\rho =w$, with w close
to $-1$.

A result obtained is that the energy density $\rho $ is a decreasing
function, for all times, when the second viscosity coefficient $\tau _{\Pi }$
is negative. It is worth recalling that the purpose of the term of TEM
associated with this coefficient is to correct the defects of the non causal
theory of relativistic fluids \cite{Landau F}, in which fluid disturbances
are propagated at superluminal speeds. This term (as it can easily be
verified from calculations ref. \cite{Romatschke IJMP}) leads to a bound for
the propagation velocity. This behavior is not affected by the change of
sign in $\tau _{\Pi }$.

In particular, a detailed study of the case where the state equation is
constant ($w=-1-\lambda $) was performed. It was found that, in order to
meet the imposed conditions, in particular $a(t)$ as an increasing function
of time, assuming one of the two following restrictions was necessary: a) $%
\tau _{\Pi }$ $<0$ and $\lambda >0$ , or b) $\tau _{\Pi }$ $>0$ and $\lambda
<0$.

In the first case, it is necessary to add the condition $\zeta \left\vert
\tau _{\Pi }\right\vert >\frac{2}{3}$, which leads to $a\left( t\rightarrow
\infty \right) \simeq \left( 2t_{P}\right) ^{-\zeta /2\lambda }$. So, the
value of the scale factor is finite but as large as we want, provided that
we make $\lambda $ small enough. It is interesting to note that the latter
requirement makes $w$ to be very close to $-1$. Moreover, $\rho \left(
t\rightarrow \infty \right) =0$ . Under these conditions the universe is
accelerated until a certain time $t_{c}$, which can be as large as we want,
for a $\lambda $ close enough to zero. After that time, the universe begins
to slow the velocity of expansion.

In the second case, $a\left( t\rightarrow \infty \right) =\infty $ was
obtained, but with $\rho \left( t\rightarrow \infty \right) =3(\zeta
/\lambda )^{2}$, which can give us a density that increases with time if $%
\left\vert \lambda \right\vert $ $<\sqrt{3}\zeta /\rho _{P}^{1/2}$. When it
is taken into account that $\rho _{P}$ is very large ($\sim 10^{100}g/m^{3}$%
), this bound is extremely small. Whenever we consider small $\lambda $, but
not to the above value, the energy density is a decreasing function of time.
Moreover, the universe is accelerated for all times provided it is $%
\left\vert \lambda \right\vert $ $<2/3+\zeta \tau _{\Pi }$, which is not a
strong constraint because it leaves open an interval of physically
reasonable values. In addition, $w$ will be limited, since $\alpha =1$, and $%
\lambda $ could be set to a value small enough to make the deviation from $%
-1 $ to fall into the observation error. Finally, an argument in favor of
this model is the fact that entropy increases for any time.

It is worth noting that, in the case previously resolved, we have started
with a linear EoS in $\rho $ and a EMT representing a viscous fluid with
causal correction. However, this can also be seen as an ideal fluid source
but with an effective EoS with $w^{\dag }$ including the viscous correction (%
$w^{\dag }$ $=p^{\dag }/\rho $, with $p^{\dag }$ given by Eq. (\ref{2.11})),
in this way it can be considered as a case particular of the inhomogeneous
EoS proposed in ref. \cite{NoOdi2005}. But this choice has not been
arbitrary but comes from the viscous fluid model with causal correction
first developed by Israel-Stewart \cite{Israel-Stewart} and later
generalized, so as to consider non-linear terms in the velocity gradients 
\cite{Romatschke IJMP}, \cite{Romatschke CQG}.

This work opens different lines for future research. On the one hand, to
consider the causal relativistic theory of fluid in a more complete version,
including nonlinear terms in the velocity gradients. This will require more
computational effort, but it would be interesting to discover the new
conditions on the more complete set of coefficients and to evaluate their
influence on dynamics of the universe. On the other hand, it can be studied
whether, at least at the level of approximation used in this work, it is
possible to establish a correspondence with some formulation from an
analysis at a more fundamental level, i.e. with a field theory formulation.
It would also be interesting to consider the state equations used in recent
papers \cite{Chavanis} to describe halos of dark matter, which are inferred
from some experiments \cite{Donato}.

\begin{acknowledgement}
I would like to thank Olimpia Lombardi for her critically reading and her
comments on the manuscript.This research was supported by the University of
Buenos Aires grant: UBACyT-01/Q710.
\end{acknowledgement}

\end{document}